\documentclass[final,5p,twocolumn]{elsarticle}

\usepackage[T1]{fontenc}
\usepackage[english]{babel}

\usepackage{hyperref}

\usepackage{amsmath}
\usepackage{amsfonts,amssymb}

\usepackage[T1]{fontenc}
\usepackage{bm}

\usepackage{graphicx}
\usepackage{lscape}
\usepackage{subfig}
\usepackage{float,caption}
\usepackage{mathtools}

\usepackage{enumitem}

\makeatletter
\def\ps@pprintTitle{%
  \let\@oddhead\@empty
  \let\@evenhead\@empty
  \def\@oddfoot{\reset@font\hfil\thepage\hfil}
  \let\@evenfoot\@oddfoot
}
\newcommand*{\smallrel}[2][.8]{%
  \mathrel{\mathpalette{\smallrel@{#1}}{#2}}%
}
\newcommand*{\smallrel@}[3]{%
  \sbox0{$#2\vcenter{}$}%
  \dimen@=\ht0 %
  \raise\dimen@\hbox{%
    \scalebox{#1}{%
      \raise-\dimen@\hbox{$#2#3\m@th$}%
    }%
  }%
}
\makeatother

\newcommand{\D}{\ensuremath \text{d}}

\begin{document}
\begin{frontmatter}

\title{\textbf{Lepton pair production in muon-nucleus scattering}}

\author[a]{G. Abbiendi}
\ead{giovanni.abbiendi@bo.infn.it}

\author[b,c]{E. Budassi}
\ead{ettore.budassi01@universitadipavia.it}

\author[c]{C. M. Carloni Calame}
\ead{carlo.carloni.calame@pv.infn.it}

\author[b,c]{A. Gurgone}
\ead{andrea.gurgone01@universitadipavia.it}

\author[c]{F. Piccinini}
\ead{fulvio.piccinini@pv.infn.it}

\affiliation[a]{organization={INFN, Sezione di Bologna},
            addressline={Viale C. Berti Pichat 6/2},
            postcode={40127},
            city={Bologna},
            country={Italy}}

\affiliation[b]{organization={Dipartimento di Fisica, \unexpanded{Università} di Pavia},
            addressline={Via A. Bassi 6},
            postcode={27100},
            city={Pavia},
            country={Italy}}

\affiliation[c]{organization={INFN, Sezione di Pavia},
            addressline={Via A. Bassi 6},
            postcode={27100},
            city={Pavia},
            country={Italy}}
            
\begin{abstract}
The MUonE experiment aims at providing a novel determination of the leading hadronic contribution to the muon anomalous magnetic moment through the study of elastic muon-electron scattering.
Since the initial-state electrons are bound in a low-$Z$ atomic target, the interaction between the incoming muons and the nuclei is expected to be the main source of experimental background.
In this article, we study the production of a real lepton pair from the muon-nucleus scattering, discussing its numerical impact in the MUonE kinematic configuration.
The process is described as a scattering of a muon in an external Coulomb field with the addition of a form factor to describe the nuclear charge distribution.
The calculation is implemented in the fully differential Monte Carlo event generator \textsc{Mesmer}, without introducing any approximation on the angular variables.
\end{abstract}


\end{frontmatter}

%=========================================%

\section{Introduction}
\label{sec:intro}

The muon magnetic moment anomaly $a_\mu = (g_\mu-2)/2$ is a fundamental observable in particle physics. 
The Muon $g-2$ experiment at FNAL has recently measured it with a precision of 0.20 ppm~\cite{Muong-2:2021ojo,Muong-2:2023cdq}.
The combination of this result with the previous BNL determination~\cite{Muong-2:2006rrc} shows a deviation of $5.0\hspace{1pt}\sigma$ from the Standard Model (SM) prediction, as compiled by the Muon $g-2$ Theory Initiative in 2020~\cite{Aoyama:2020ynm}.

The theoretical uncertainty of $a_\mu$ is dominated by the Hadronic Vacuum Polarisation (HVP) contribution $a_\mu^{\text{HVP}}$, which is traditionally evaluated through the time-like dispersive approach~\cite{Keshavarzi:2019abf,Davier:2019can,Benayoun:2019zwh}, based on the measurement of the hadronic production in $e^+e^-$ collisions.
However, the accuracy of this method is limited by the existence of several resonances in the low-energy cross section.
Furthermore, the CMD-3 collaboration has recently published a new measurement of the ${e^+e^- \to \pi^+\pi^-}$ cross section~\cite{CMD-3:2023alj,CMD-3:2023rfe}, which disagrees with the results of previous experiments~\cite{BaBar:2012bdw,BESIII:2015equ,KLOE-2:2017fda}.
In addition, the data-driven determinations of $a_\mu^{\text{HVP}}$ disagree with the BMW lattice prediction~\cite{Borsanyi:2020mff}, which would significantly reduce the muon $g-2$ anomaly.
All these tensions suggest possible inaccuracies in the SM prediction of $a_\mu$, preventing a reliable comparison with the experimental value.

An alternative data-driven approach to evaluate $a_\mu^{\text{HVP}}$ has been proposed in~\cite{CarloniCalame:2015obs}. 
The new method is based on the measurement of the hadronic contribution to the running of the QED coupling $\Delta \alpha(t)$ in the space-like region $t<0$, where the HVP contribution is a smooth function~\cite{Lautrup:1971jf,Balzani:2021del}.
The function $\Delta\alpha(t)$ can be obtained directly from the cross section of a $t$-channel process, such as the elastic muon-electron scattering~\cite{Abbiendi:2016xup}.
In this regard, the MUonE experiment has been proposed to study the elastic scattering of a muon or antimuon beam of 160~GeV on atomic electrons~\cite{MUonE:LoI}.
After the proposal, there has been a considerable progress in the development of detectors and data analysis strategies~\cite{Abbiendi:2019qtw,Abbiendi:2021xsh,Abbiendi:2022oks,NunzioPilato:2024fgk,Magherini:2024mav,Spedicato:2023,Ignatov:2023wma}, as well as in the investigation of possible new physics contamination~\cite{Masiero:2020vxk,Dev:2020drf,Atkinson:2022qnl,Le:2023ceg}.
The MUonE experiment will provide an independent determination of $a_\mu^{\text{HVP}}$ with an uncertainty below the percent level, comparable with the time-like and lattice results.

The MUonE precision goal requires a theoretical computation of ${\mu^\pm e^- \to \mu^\pm e^-}$ with an accuracy of 10~ppm on the shape of differential cross sections~\cite{Banerjee:2020tdt}.
In order to achieve such a precision, the calculation must include QED corrections at least at the next-to-next-to-leading order~\cite{Alacevich:2018vez,CarloniCalame:2020yoz,Budassi:2021twh,Broggio:2022htr,Mastrolia:2017pfy,DiVita:2018nnh,Bonciani:2021okt,Fael:2022rgm,Ahmed:2023htp,Badger:2023xtl,Engel:2023ifn,Engel:2023rxp} with the addition of hadronic contributions~\cite{Fael:2018dmz,Fael:2019nsf}.
The computation must be implemented in a fully differential Monte Carlo~(MC) event generator in order to be used in the MUonE data analysis.
In this regard, two independent MC codes, \textsc{Mesmer}~\cite{Alacevich:2018vez,CarloniCalame:2020yoz,Budassi:2021twh} and \textsc{McMule}~\cite{Banerjee:2020rww,Broggio:2022htr}, are currently under development.

In addition, the impact of the possible background processes must be precisely evaluated.
In this regard, the \textsc{Mesmer} event generator has been extended to include the simulation of ${\mu^\pm e^- \to \mu^\pm e^- \ell^+\ell^-}$ with~$\ell=\{e,\mu\}$~\cite{Budassi:2021twh} and ${\mu^\pm e^- \to \mu^\pm e^- \pi^0}$~\cite{Budassi:2022kqs}.
However, since the initial-state electrons are bound in a low-$Z$ atomic target (beryllium-9 or carbon-12), the main source of experimental background is given by the muon-nucleus scattering. 
In fact, its cross section scales with $Z^2$, while the muon-electron scattering scales with $Z$. 

In this article, we focus on the production of a real lepton pair from the scattering of an incoming muon on a nucleus at rest, namely ${\mu^\pm X \to \mu^\pm X \ell^+ \ell^-}$, where $X$ denotes a nucleus.
This process is particularly important because it can resemble an elastic event if one of the final leptons is not detected.
Furthermore, it can be a relevant background in possible new physics searches with the MUonE apparatus~\cite{Asai:2021wzx,Galon:2022xcl,GrillidiCortona:2022kbq}, especially in the case of new light mediators decaying in a $e^+e^-$ pair.

The process ${\mu^\pm X \to \mu^\pm X e^+ e^-}$ is already implemented in the \textsc{Geant4} toolkit~\cite{GEANT4:2002zbu,Bogdanov:2006kr}, following the calculations presented in~\cite{Kokoulin:1970,Kokoulin:1971}.
In this regard, the MUonESim application has been used to compare the implementation of the process in different \textsc{Geant4} versions~\cite{Asenov:2022uds}.
In particular, while the angular distribution of the $e^+e^-$ pair has been refined in the update 10.7, the muon scattering angle is neglected in all available versions.
This is not a valid approximation for MUonE, where the muon scattering angle is expected to be of about ${\cal O}(m_\mu / E_\mu) \sim 0.66$~mrad, in contrast to a detector resolution of ${20-30}$~$\mu$rad~\cite{MUonE:LoI}.
Since the QED dynamics enhances the region 
of very small muon scattering angles, a new fully differential calculation for the process ${\mu^\pm X \to \mu^\pm X \ell^+ \ell^-}$, without any approximation on the angular variables, is required.

In order to achieve this goal, we start in Section~\ref{sec:calc} presenting a new calculation of the tree-level differential cross section for the process ${\mu^\pm X \to \mu^\pm X \ell^+ \ell^-}$ and its implementation in the \textsc{Mesmer} event generator.
In Section~\ref{sec:num_res}, we exhibit a collection of numerical results which are relevant for the MUonE experiment.
Finally, our study is summarised in Section~\ref{sec:concl}.

%=========================================%

\section{The calculation}
\label{sec:calc}

The process described in this article is
\begin{equation}
\mu^\pm(p_1) \, X \to \mu^\pm(p_2) \,X \, \ell^+(p_3) \, \ell^-(p_4) \,, 
\label{eq:process}
\end{equation}
where $p_i$ denotes the four-momentum of the corresponding lepton, $X$ represents any nuclear isotope, and $\ell=\{e,\, \mu\}$.
We do not consider the production of a $\tau^+\tau^-$ pair, because its contribution is negligible for the MUonE kinematics.

The process is described as a scattering of a muon in an external Coulomb field, generated by a nucleus at rest.
Since the nuclear recoil is neglected, the energy is conserved in the process, while the three-momentum is not.
The nuclear charge distribution is described by using the form factors available in literature~\cite{DeVries:1987atn,Fricke:1995zz}.
As will be discussed in Section~\ref{sec:num_res}, this approach is supported by the MUonE acceptance, which is limited to small scattering angles, i.e. small transferred momenta.

\begin{figure}[t]
    \centering
    \includegraphics[width = 0.49\columnwidth]{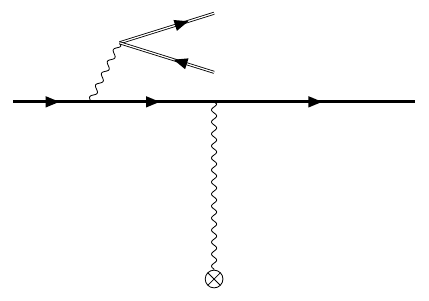}
    \includegraphics[width = 0.49\columnwidth]{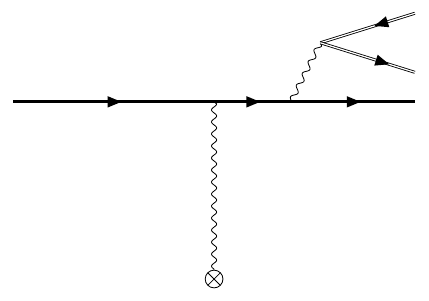}
    \includegraphics[width = 0.49\columnwidth]{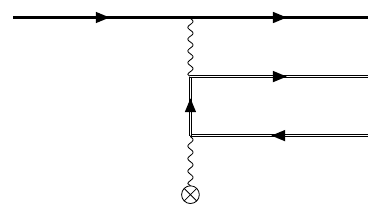}
    \includegraphics[width = 0.49\columnwidth]{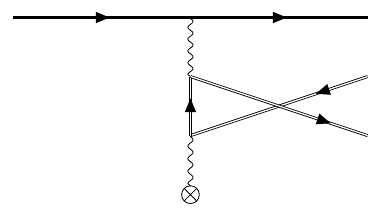}
    \caption{The four tree-level diagrams contributing to the process. The lepton pair can be emitted either from the muon propagator (radiative diagrams) or from the virtual photon exchanged with the nuclear field (peripheral diagrams). For the production of a $\mu^+\mu^-$ pair, the four additional exchange diagrams must be included as well.}
    \label{fig:diagrams}
\end{figure}

The lepton pair can be emitted either from the incident muon through photon radiation or from the virtual photon exchanged with the nucleus.
The corresponding diagrams are depicted in Fig.~\ref{fig:diagrams}.
The overall sign of the interference terms between the two diagram topologies depends on the charge of the incoming muon.
The cross section is therefore different for positive and negative incoming muons, although the difference is negligible for high-energy beams.

The matrix elements have been calculated independently with \textsc{FORM}~\cite{Vermaseren:2000nd,Kuipers:2012rf,Ruijl:2017dtg} and \textsc{Package-X}~\cite{Patel:2015tea,Patel:2016fam}, without neglecting any mass.
At this stage, the calculation is performed assuming a Coulomb field generated by a point-like charge $Ze$.

In order to account for the finite size of the nucleus, the matrix element can be multiplied with a form factor $F(q)$, where $q$ is the transferred three-momentum.
The form factor is related to the nuclear charge distribution $\rho(r)$ via the Fourier transformation 
\begin{equation}
    F(q)=\frac{4\pi}{Ze}\int_0^\infty \D rr^2\rho(r)\frac{\sin(qr)}{qr} \,,
    \label{eq:nff}
\end{equation}
where the charge distribution is normalised so that
\begin{equation}
    4\pi \int_0^\infty \D rr^2\rho(r) = Ze \,.
\end{equation}
The functional form of $F(q)$ for a given isotope depends on the parametrisation chosen for $\rho(r)$.
The form factor for a point-like nucleus is simply $F=1$.
More realistic descriptions rely on model assumptions and experimental data, both from atomic spectroscopy and scattering experiments~\cite{DeVries:1987atn,Fricke:1995zz}.
In order to estimate the corresponding uncertainty, we use different models for the same isotope, following the same approach as in~\cite{Heeck:2021adh,Heeck:2022wer}.
The models used in this work are:
\begin{itemize}
\item The one parameter Fermi model (1pF)
\begin{equation}
\rho(r) = \frac{\rho_0}{1+\exp{\frac{r-c}{z}}} \,,
\label{eq:1pF}
\end{equation}
where $\rho_0$ is a normalisation constant, $z \simeq 0.52$~fm is the surface thickness, and $c$ is the half-density radius, which can be evaluated from the experimental value of the root-mean-square charge radius. 

\item The modified harmonic oscillator model (MHO)
\begin{equation}
\rho(r) = \rho_0' \left( 1+\omega\frac{r^2}{a^2}\right) \exp{\left(-\frac{r^2}{a^2}\right)} \,,
\end{equation}
where $\rho_0'$ is a normalisation constant, while $a$ and $\omega$ are two input parameters that can be extracted from experimental data.

\item The Fourier-Bessel expansion (FB)
\begin{equation}
\rho(r) = 
\begin{cases}
\;\sum\limits_{k=1}^n a_k\, j_0\hspace{-1pt}\left( \frac{k\pi r}{R}\right) \quad &r \leq R \\[8pt]
\;0 &r > R\,,
\label{eq:FB}
\end{cases}
\end{equation}
where $j_0(x) = \sin(x)/x$ is the 0-th order spherical Bessel function, $R$ is an input parameter, and $a_k$ are the expansion coefficients, normalised so that
\begin{equation}
\sum\limits_{k=1}^n a_k \frac{(-1)^{k+1}}{k^2} = Ze \frac{\pi}{4R^3} \,.
\end{equation}
\end{itemize}
In the following, we focus on beryllium-9 and carbon-12, which are the two materials proposed for the MUonE target.
In this regard, we consider the 1pF model for both isotopes, using the nuclear radii listed in~\cite{2013ADNDT..99...69A}.
In addition, we consider an alternative parametrisation for each material: the MHO model for beryllium-9 and the FB expansion for carbon-12, using the results reported in~\cite{Jansen:1972iui,Offermann:1991ft}.
As discussed in~\cite{Heeck:2021adh,Heeck:2022wer}, the lack of extensive results on nuclear form factors does not allow a more detailed comparison between different models and experiments.
In particular, the simple 1pF model is the only model that presents results for both beryllium-9 and carbon-12.

Assuming only energy conservation, the phase space of the process is given by 
\begin{equation}
    \D \Phi_3 = \prod_{i=2}^4\frac{\D ^3\vec{p}_i}{(2\pi)^32E_i}\delta^1\left(E_1-\sum_{j=2}^4E_j\right) ,
\end{equation}
where $E_i$ denotes the energy of the particle with four-momentum $p_i$.
In order to implement an efficient MC integration and event generation, we decompose the phase space in four different channels, corresponding to the four diagrams of Fig.~\ref{fig:diagrams}.
Assuming small scattering angles, we use a different importance sampling for each channel, inspired by the propagator structure of the corresponding diagram.
Accordingly, the phase space is parametrised and generated as follows:
\begin{enumerate}
\item Initial-state radiation: 
\begin{align}
\begin{split}
     \D\Phi_3\: &=\: \frac{1}{4 (2\pi)^2}\, \D Q^2 \, \D E_Q \,  \D\hspace{-1pt}\cos\vartheta_Q  
     \D\phi_Q \\
     &\times\:  \D\hspace{-1pt}\cos\vartheta_2 \,  \D \phi_2 \,  \D \Phi_2 (Q\to p_3+p_4) \\
     &\times\: E_Q \vert \vec{Q} \vert \,,
\end{split}
\end{align}
where $Q^2=(p_3+p_4)^2$ is the squared invariant mass of the emitted leptonic pair, $\vartheta_Q$ and $\phi_Q$ are the polar and azimuthal angles of the leptonic pair momentum $\vec{Q}$  
and $\D\Phi_2 (Q\to p_3+p_4)$ is the Lorentz-invariant phase space of the 2-body decay $Q\to p_3+p_4$.
In this channel, the laboratory frame is chosen with the $z$-axis directed along the incoming muon.
The dilepton invariant mass $Q^2$ is sampled according to $1/Q^2$, while the polar angle $\vartheta_2$ is sampled according to $\vert \vec{q\hspace{1pt}} \vert^{-4}$, where $q = p_1 - p_2 - p_3 - p_4$ is the momentum exchanged with the nucleus.
The momenta $p_3$ and $p_4$ are generated in the pair rest frame, where $\vec{Q} = 0$, and boosted back in the laboratory frame.
The muon energy $E_2$ is given by the energy conservation, while the azimuthal angle $\phi_2$ is sampled uniformly.
However, since the system has a cylindrical symmetry around the $z$-axis, all momenta are generated imposing $\phi_2=0$, and then rotated by $\phi_2$.
\item Final-state radiation: 
\begin{align}
\begin{split}
     \D \Phi_3\: &=\: \frac{1}{4 (2\pi)^2}\, \D Q^2 \, \D E_Q \,  \D\hspace{-1pt}\cos\vartheta_Q \D \phi_Q \\
     &\times\: \D\hspace{-1pt}\cos\vartheta_2 \,  \D \phi_2 \,  \D \Phi_2 (Q\to p_3+p_4) \\
     &\times\: E_Q \vert \vec{Q} \vert \,.
\end{split}
\end{align}
In this channel, the reference frame is chosen such that the $z$-axis is parallel to $\vec{p}_2$. 
After the generation of $\vartheta_2$, the two angles $\vartheta_Q$ and $\phi_Q$ are calculated in the laboratory frame assuming $\phi_2=0$. 
The remaining generation steps of $p_3$ and $p_4$, as well as the overall rotation by $\phi_2$, follow the same reasoning of channel~1.
Again, the function $1/Q^2$ is used to sample $Q^2$, while $\vert \vec{q\hspace{1pt}} \vert^{-4}$ is used to sample $\vartheta_2$.
The other variables are sampled uniformly in the respective reference frames.
\item Peripheral diagram:
\begin{align}
\begin{split}
     \D \Phi_3\: &=\: \frac{1}{8 (2\pi)^9} \D \vert \vec{p}_2 \vert \, \D \vert \vec{p}_3 \vert \\
     &\times\: \D \phi_2 \,  \D \phi_3 \,  \D \phi_4 \\
     &\times\: \D\hspace{-1pt}\cos\vartheta_2\, \D\hspace{-1pt}\cos\vartheta_3 \, \D\hspace{-1pt}\cos{\vartheta_4} \\
     &\times\: \frac{\vert \vec{p}_2\vert^2}{E_2}  \, \frac{\vert \vec{p}_3\vert^2}{E_3} \\
     &\times\: \sqrt{(E - E_2 - E_3)^2 - m_\ell^2}\,,
\end{split}
\end{align}
where $E$ is the muon beam energy and $m_\ell$ is the pair lepton mass. 
As for the previous channels, the momenta are generated in the laboratory frame with $\phi_2=0$ and then rotated by $\phi_2$.
The angle $\vartheta_2$ is sampled according to $[(p_1-p_2)^2]^{-1}$, $\vartheta_3$ according to $[(p_1-p_2-p_3)^2 - m_\ell^2]^{-1}$, and $\vartheta_4$ according to $\vert \vec{q\hspace{1pt}} \vert^{-4}$.
\item Crossed peripheral diagram: 
the same as the previous channel with the inversion of $\vartheta_3$ and $\vartheta_4$.
In particular, the function $[(p_1-p_2-p_4)^2 - m_\ell^2]^{-1}$ is used to sample $\vartheta_4$, while $\vert \vec{q\hspace{1pt}} \vert^{-4}$ is used to sample $\vartheta_3$.
\end{enumerate}

The process has been implemented in the fully differential MC event generator \textsc{Mesmer}\footnote{An up-to-date version of \textsc{Mesmer} can be found at \url{https://github.com/cm-cc/mesmer}.} in order to be used for numerical simulations and experimental studies.

%=========================================%

\section{Numerical results}
\label{sec:num_res}

In this section, we study the numerical impact of the process ${\mu^\pm X \to \mu^\pm X \ell^+ \ell^-}$ for the typical running conditions and event selections of the MUonE experiment.
The kinematic cuts used in the simulation are defined in Section~\ref{subsec:event_selection}.
In Section~\ref{subsec:num-tot}, we report the results for total cross section, including a comparison with the \textsc{Geant4} implementation of the same process for validation purposes.
In Section~\ref{subsec:num-basic}, we present the results for the more relevant differential cross sections.
The impact of the nuclear form factor is discussed in Section~\ref{subsec:num-ff}.
Finally, in Section~\ref{subsec:num-aco-el-cuts}, we discuss the impact of possible additional cuts, defined to increase the rejection of this process.

\subsection{Event selection criteria}
\label{subsec:event_selection}

In order to evaluate the numerical impact of the process ${\mu^\pm X \to \mu^\pm X \ell^+ \ell^-}$ for the MUonE experiment, we define a set of fiducial cuts inspired by the experimental acceptances and the event selection criteria.
In the following, the muon scattering angle is denoted with $\vartheta_\mu$, the electron (or positron) scattering angle with $\vartheta_e$, the muon energy with $E_\mu$, and the electron (or positron) energy with $E_e$.
All quantities are measured in the laboratory frame with the $z$-axis directed along the incoming muon, whose 
energy is fixed to $E=160$~GeV. 

The acceptance region of the final particles is defined as in the previous studies on the MUonE experiment with the \textsc{Mesmer} event generator~\cite{Alacevich:2018vez,CarloniCalame:2020yoz,Budassi:2021twh,Budassi:2022kqs}:
\begin{itemize}
    \item \emph{Basic acceptance cuts}: $0.2$~mrad~$< \vartheta_\mu < 4.84$~mrad, $E_\mu > 10.23$~GeV, $\vartheta_e<32$~mrad, $E_e> 0.2$~GeV.  
\end{itemize} 
The above cuts represent a simplified experimental event selection which should 
be properly defined through detector simulation and track reconstruction.
Since the process ${\mu^\pm X \to \mu^\pm X \ell^+ \ell^-}$ is enhanced for small scattering angles, the cut on the minimum muon angle has the effect of suppressing most of its contribution.
Furthermore, the cut $\vartheta_e<32$~mrad implies $\vartheta_\mu > 0.2$~mrad for ${\mu^\pm e^- \to \mu^\pm e^-}$ events.
Hence, this cut is crucial to select the elastic signal, while rejecting most of the background events.

In the case $\ell = e$, since the MUonE experiment cannot distinguish an electron from a positron, the cuts on $\vartheta_e$ and $E_e$ are applied to both particles.
In the case $\ell = \mu$, we assume that only one muon is correctly identified as a muon, while the other reconstructed muons are erroneously identified as electrons. 
This choice is due to the fact that an event with more than one muon would be rejected in the experimental analysis.
If only one muon satisfies the cuts on $\vartheta_\mu$ and $E_\mu$, it is identified as the only muon in the process, while the other muons are considered as electrons and subjected to the cuts on $\vartheta_e$ and $E_e$ as in the case $\ell = e$.
If multiple muons satisfy the cuts on $\vartheta_\mu$ and $E_\mu$, the one identified as a muon is chosen randomly, while the others are considered as electrons.

All events are classified according to the number of detected particles: two-track events (one muon and one electron) and three-track events (one muon and two electrons).
In the case of two-track events, the muon satisfies the cut on $\vartheta_\mu$ and $E_\mu$, but only one of the two electrons meets the cut on $\vartheta_e$ and $E_e$.  
This is the most important case, because the event has the same signature of the elastic muon-electron scattering.
In the case of three-track events, all particles satisfy the respective cuts.
In addition, if the relative angle between two tracks is less than the angular resolution, i.e. ${\delta\vartheta < 20~\mu}$rad, the number of tracks is reduced by one.
This takes into account the fact that the experiment cannot distinguish two overlapping tracks as separate.

An additional cut on the maximum three-momentum exchanged with the nucleus is required to ensure the validity of the external field approximation:
\begin{itemize}
    \item \emph{Maximum transferred momentum}: ${|q|^2<0.6}$~GeV$^2$.
    The chosen threshold value corresponds to the three-momentum exchanged in Mott scattering $(\mu X \to \mu X)$ for ${E=160}$~GeV and ${\theta_\mu = 4.84}$~mrad. 
\end{itemize} 
As we show in Section~\ref{subsec:num-ff}, the impact of this cut is negligible, supporting the model adopted for the calculation.
In the following, the use of this cut is always implied.

We define two additional cuts to increase the rejection of the lepton pair background without significantly reducing the elastic signal cross section:

\begin{itemize}
    \item \emph{Acoplanarity cut}: $\xi = \vert \pi - \vert \phi_e - \phi_\mu \vert \vert < \xi_c$, where $\phi_{e}$ and $\phi_{\mu}$ are the electron and muon azimuthal angles. 
    We choose, for illustration purposes, the value ${\xi_c = 400}$~mrad, which appears to be more realistic than the value $\xi_c = 3.5$~mrad used in~\cite{Budassi:2021twh}. 
    \item \emph{Elasticity cut}: $\delta < 0.2$ mrad, where $\delta$ is the minimal Cartesian distance between the point $(\vartheta_e,\vartheta_\mu)$ and the elasticity curve~\cite{Budassi:2021twh}
    \begin{equation}
    \theta_\mu(\theta_e) = \arctan{\left[\frac{2 m_e r \cos\theta_e\sin\theta_e}{E-r(rE+2m_e)\cos^2\theta_e}\right]}\,,
    \end{equation}
    where $r=\left(E^2-m_\mu^2\right)^{\frac12}\!/\left(E+m_e\right)$.
\end{itemize}

Neglecting the emission of radiation, for the elastic muon-electron scattering we have $\xi = 0$ and $\delta = 0$.
Since the three-track events can always be rejected using the number of tracks, we apply these last two cuts only to two-track events, which are more difficult to distinguish from the elastic signal.

\subsection{Total cross section}
\label{subsec:num-tot}

As a validation of the code, we compare the \textsc{Mesmer} predictions with the outcome of the \textsc{Geant4} process \texttt{G4MuPairProduction}~\cite{Bogdanov:2006kr}.
For this purpose, we consider a fully inclusive angular acceptance with only a cut on the minimum total energy of the emitted pair, neglecting the effect of the nuclear form factor.
Since the \textsc{Geant4} implementation includes both the scattering on the nucleus and on atomic electrons, the latter contribution is removed by multiplying the result by $Z/(Z+\zeta)$, where $\zeta=\zeta(Z,E)$ is a parameter defined in~\cite{Geant4:Manual} accounting for the electronic binding effects.
In our case, for $Z=6$ and $E=160$~GeV, we have $\zeta \approx 0.967$.
The result for the process ${\mu^+ \,\text{C} \to \mu^+ \,\text{C}\: e^+ e^-}$ for different energy thresholds is reported in Tab.~\ref{tab:geant}.
The cross sections evaluated by the two codes show a relative difference below 5\%, which is consistent with the different model assumptions.

\begin{table}
\centering
\begin{tabular}{|c|c|c|c|}
 \hline
 $E_{ee}$ $[$GeV$]$ & \textsc{Mesmer} $[\mu$b$]$ & \textsc{Geant4} $[\mu$b$]$ & Diff.\\
 \hline
 5   & 47.700(74) & 45.986 & 3.7\% \\
 10  & 12.080(20) & 12.082 & -0.02\% \\          
 20  & 2.737(54)  & 2.766  & -1.0\% \\       
 40  & 0.5213(88) & 0.543 & -4.0\% \\          
 \hline
\end{tabular}
 \caption{Comparison between \textsc{Mesmer} and \textsc{Geant4} for the total cross section of the process ${\mu^+ \,\text{C} \to \mu^+ \,\text{C}\: e^+ e^-}$ for different cuts on the minimum total energy of the emitted pair and a fully inclusive angular acceptance.}
 \label{tab:geant}
\end{table}

As an additional cross check, the fully inclusive cross section of the process ${\mu^+ \,\text{C} \to \mu^+ \,\text{C}\: \mu^+ \mu^-}$ can be compared with the analytical expression presented in~\cite{Lee:2023cwb}.
Neglecting the form factor, the \textsc{Mesmer} result is 196.3(9)~nb, which is compatible with the analytical value of 196.74~nb.

The total cross sections for all configurations, integrated using the {\it basic acceptance cuts}, are given in Tab.~\ref{tab:xtot}. 
In order to provide a reference value, the total cross section for the Mott scattering $(\mu X \to \mu X)$ in the same kinematic region is 4.064~mb for beryllium and 9.145~mb for carbon.  

\begin{table}
\centering
\begin{tabular}{|c|c|}
 \hline
 Process & Cross section $[$nb$]$ \\
 \hline
 $\mu^+ \,\text{Be} \to \mu^+ \,\text{Be}\: e^+ e^-$   & 488.31(25) \\
 $\mu^- \,\text{Be} \to \mu^- \,\text{Be}\: e^+ e^-$   & 488.38(27) \\
 $\mu^+ \,\text{Be} \to \mu^+ \,\text{Be}\: \mu^+ \mu^-$   & 10.624(28) \\
 $\mu^- \,\text{Be} \to \mu^- \,\text{Be}\: \mu^+ \mu^-$   & 10.625(27) \\ 
 \hline
 $\mu^+ \,\text{C} \to \mu^+ \,\text{C}\: e^+ e^-$   & 1102.58(55) \\
 $\mu^- \,\text{C} \to \mu^- \,\text{C}\: e^+ e^-$   &  1103.08(63)\\
 $\mu^+ \,\text{C} \to \mu^+ \,\text{C}\: \mu^+ \mu^-$   & 24.111(55) \\
 $\mu^- \,\text{C} \to \mu^- \,\text{C}\: \mu^+ \mu^-$   & 24.106(61) \\      
 \hline
\end{tabular}
 \caption{Total cross section for the process ${\mu^\pm \,\text{X} \to \mu^\pm \,\text{X}\: \ell^+ \ell^-}$ for $X=\{\text{Be,}\,\text{C}\}$ and $\ell=\{e,\mu\}$ with \textit{basic acceptance cuts}.}
 \label{tab:xtot}
\end{table}

In the following, we focus on the scattering of a positive muon on a carbon target, which is the configuration adopted so far in the MUonE test runs~\cite{Abbiendi:2021xsh,Pilato:2023}.

\subsection{Numerical results for basic acceptance cuts}
\label{subsec:num-basic}

\begin{figure}[t]
    \centering
    \includegraphics[width = \columnwidth]{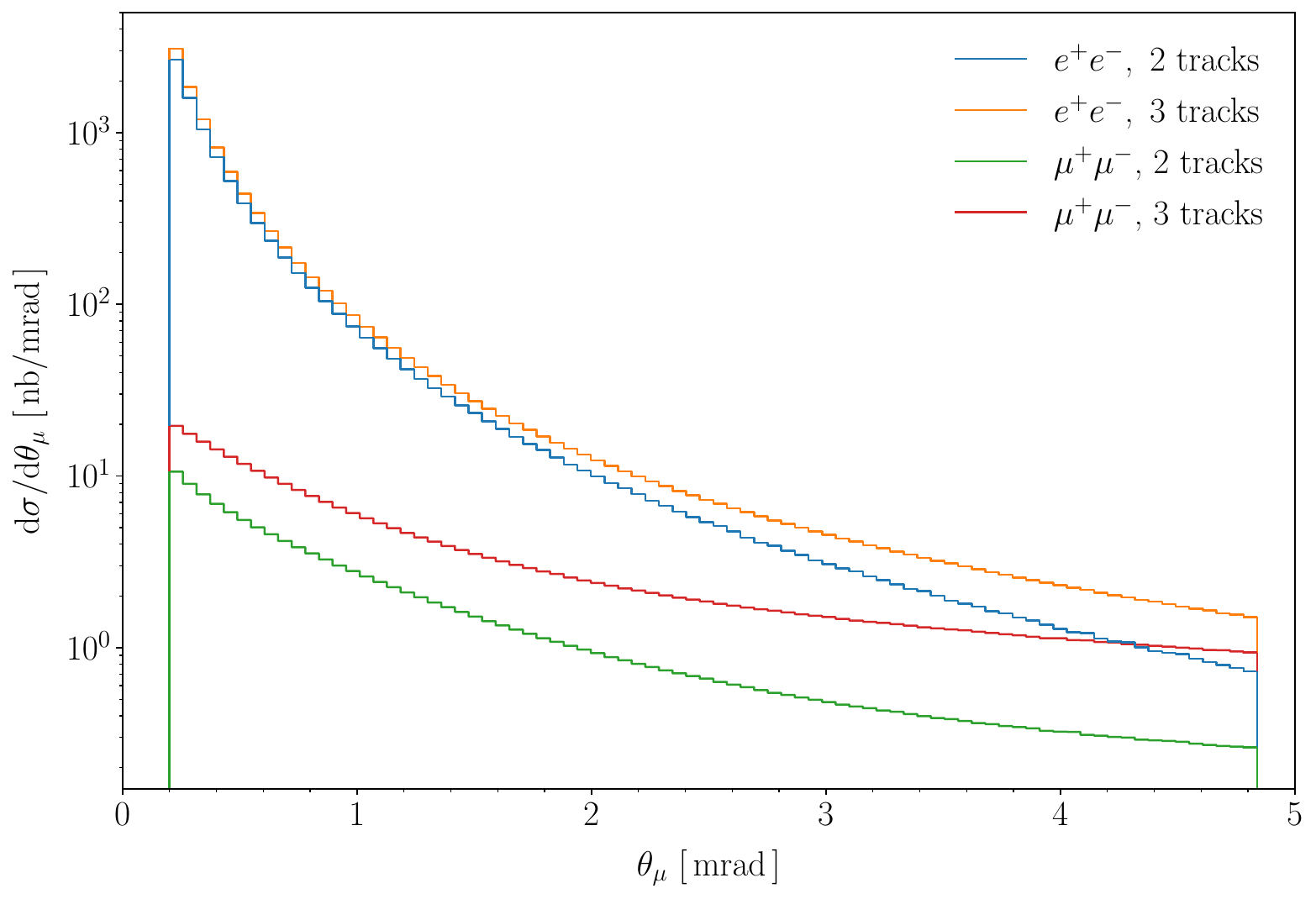}
    \caption{Differential cross section with respect to the muon angle $\vartheta_\mu$ for the production of a $e^+e^-$ or $\mu^+\mu^-$ pair.}
    \label{fig:th_mu}
\end{figure}

\begin{figure}[t]
    \centering
    \includegraphics[width = \columnwidth]{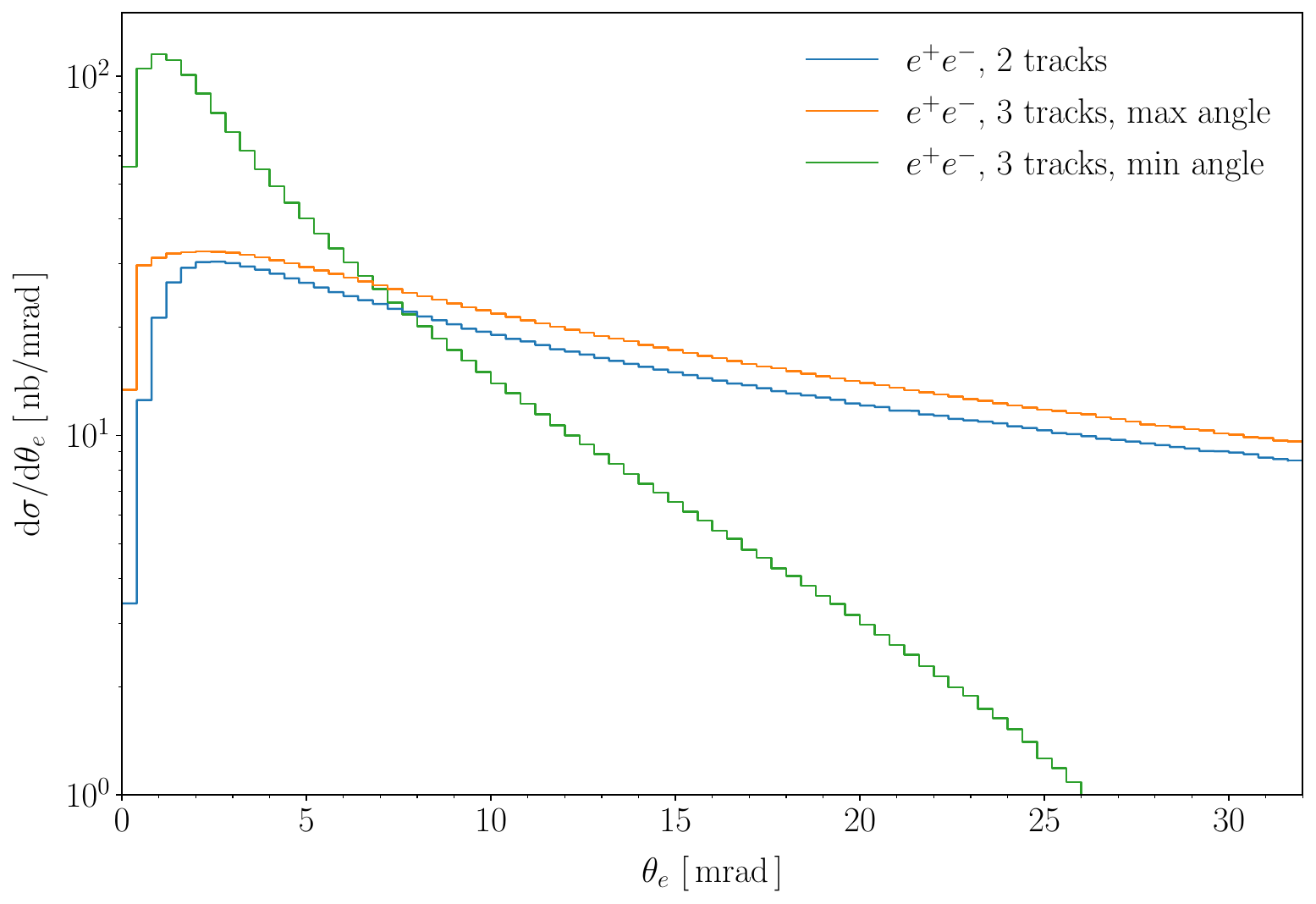}
    \caption{Differential cross section with respect to the electron angle $\vartheta_e$ for the production of a $e^+e^-$ pair.}
    \label{fig:th_e}
\end{figure}

\begin{figure}[t]
    \centering
    \includegraphics[width = \columnwidth]{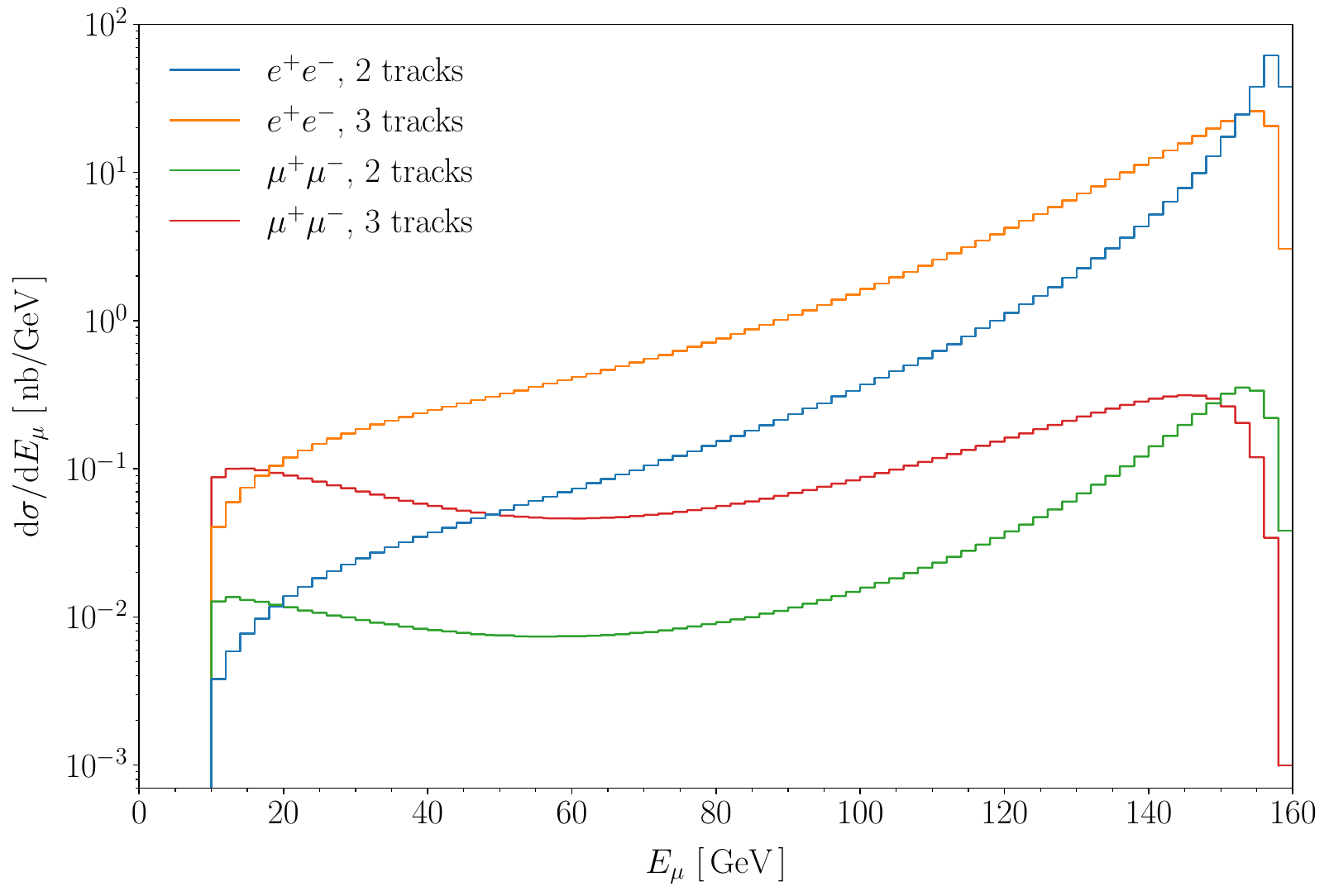}
    \caption{Differential cross section with respect to the muon energy $E_\mu$ for the production of a $e^+e^-$ or $\mu^+\mu^-$ pair.}
    \label{fig:e_mu}
\end{figure}

\begin{figure}[t]
    \centering
    \includegraphics[width = \columnwidth]{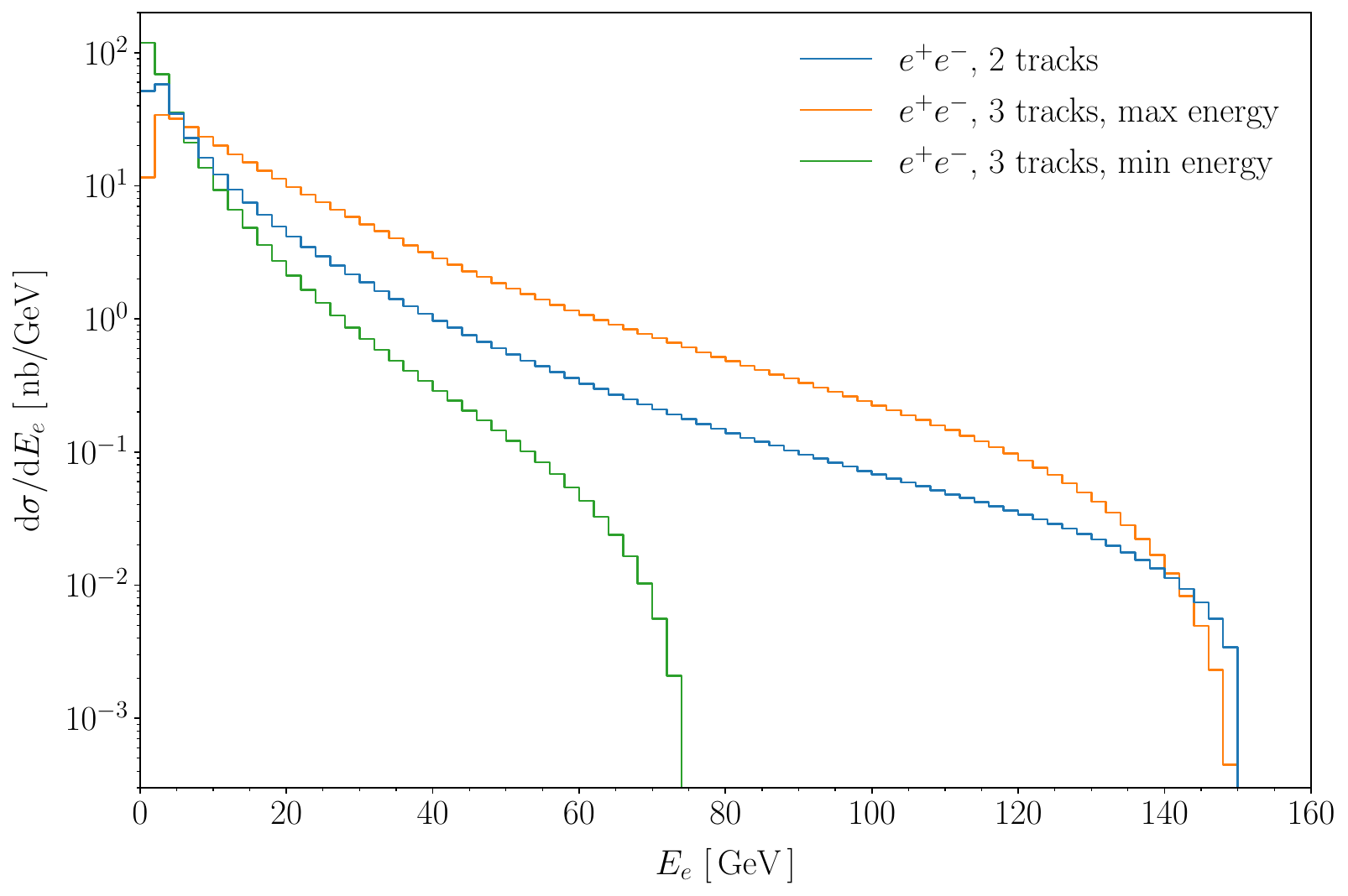}
    \caption{Differential cross section with respect to the electron energy $E_e$ for the production of a $e^+e^-$ pair.}
    \label{fig:e_e}
\end{figure}

In order to characterise the shape of the events for {\it basic acceptance cuts}, we present here the 
differential distributions of the scattering angles and the energies of the final particles, 
namely $\D \sigma / \D \vartheta_\mu$ (Fig.~\ref{fig:th_mu}), 
$\D \sigma / \D \vartheta_e$ (Fig.~\ref{fig:th_e}), 
$\D \sigma / \D E_\mu$ (Fig.~\ref{fig:e_mu}), 
and $\D \sigma / \D E_e$ (Fig.~\ref{fig:e_e}). 
In all plots we display both the case of two and three reconstructed tracks, using the 1pF model for the nuclear form factor.

Figs.~\ref{fig:th_mu} and~\ref{fig:e_mu} show that the cross section for muon pair production is, as expected, lower than the one for electron pair production, by a factor of ${\cal O}(10^2)$ in the region of small muon scattering angles or large muon energies, where the cross section for electron pairs is enhanced. 
In the region of larger muon scattering angles or smaller muon energies, the muon pair production cross section can become of the same order of magnitude as the electron pair production. 
The same figures show that, with the adopted event selection, the differential cross sections for the events with three tracks are larger than the cross sections with only two tracks, although the order of magnitude is the same.
In the case of three-track events, Figs.~\ref{fig:th_e} and~\ref{fig:e_e} display the distributions of the minimum and maximum electron scattering angle and energy, in order to discriminate between the two detected electrons. 
\begin{figure}[t]
    \centering
    \includegraphics[width = \columnwidth]{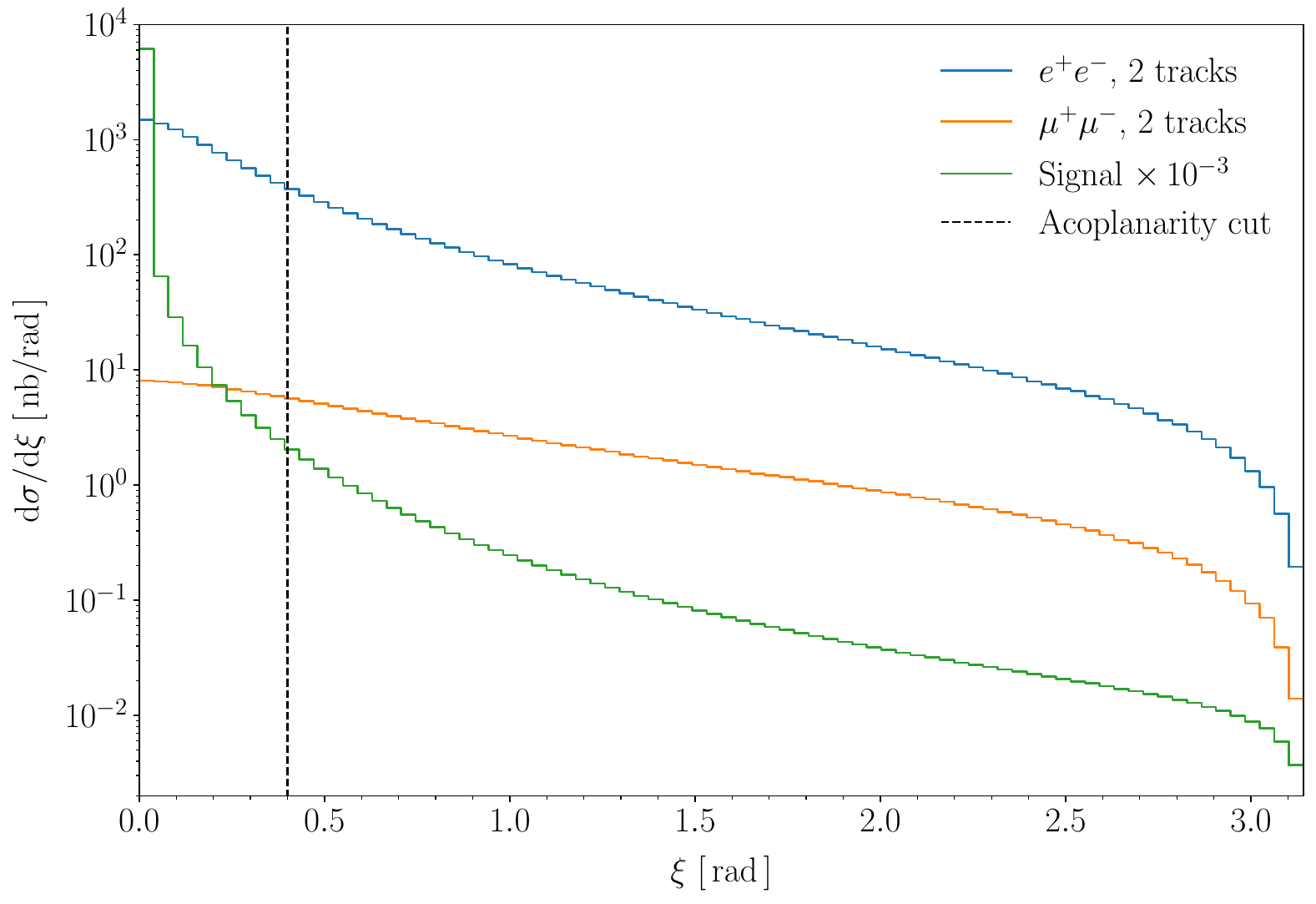}
    \caption{Differential cross section with respect to the acoplanarity parameter $\xi$ for two-tracks ($e^+ e^-$ and $\mu^+ \mu^-$) events, compared with the elastic signal one (rescaled by a factor of thousand). 
    The black dashed line represents the cut parameter $\xi_c=400$~mrad.}
    \label{fig:acoplan}
\end{figure}
Fig.~\ref{fig:acoplan} shows the distribution $\D \sigma / \D \xi$, where $\xi$ is the acoplanarity, for the events with two tracks and for both the electron and muon pair production processes, compared with the elastic signal events. The vertical black dashed line corresponds to the chosen value of the maximum acoplanarity cut of $400$~mrad. The slope of the distributions, in particular the one referring to electron pairs, suggests that a cut on this variable can be useful in reducing the background to elastic events. 
\begin{figure}[t]
    \centering
    \includegraphics[width = .995\columnwidth]{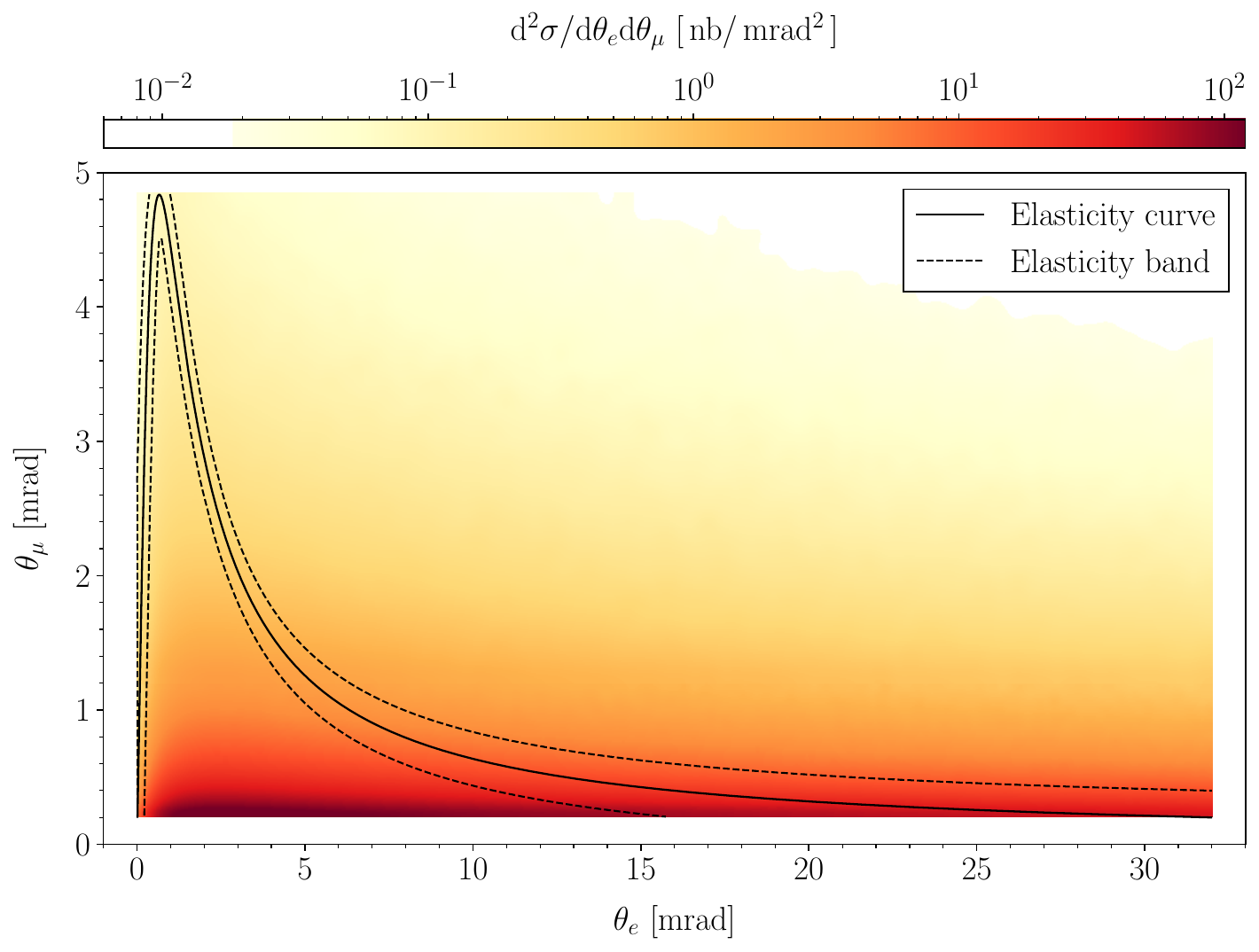}
    \caption{Double differential cross section $\D^2\sigma/\D\vartheta_e \D\vartheta_\mu$ for two-tracks events and $\ell=e$. 
    The elasticity curve is denoted by a black solid line, while the black dashed lines delimit the area where the events satisfy the elasticity cut.}
    \label{fig:2d_e}
\end{figure}
The $\vartheta_\mu - \vartheta_e$ correlation plot, with superimposed the  elasticity curve as predicted for the elastic signal events and the region allowed by the elasticity cut, is shown in Fig.~\ref{fig:2d_e}. 
It is clear that there is room for reducing the impact of the background by 
considering the additional cuts described in Section~\ref{subsec:event_selection}.

\subsection{Numerical impact of the nuclear form factor}
\label{subsec:num-ff}
\begin{figure}[t]
    \centering
    \includegraphics[width = \columnwidth]{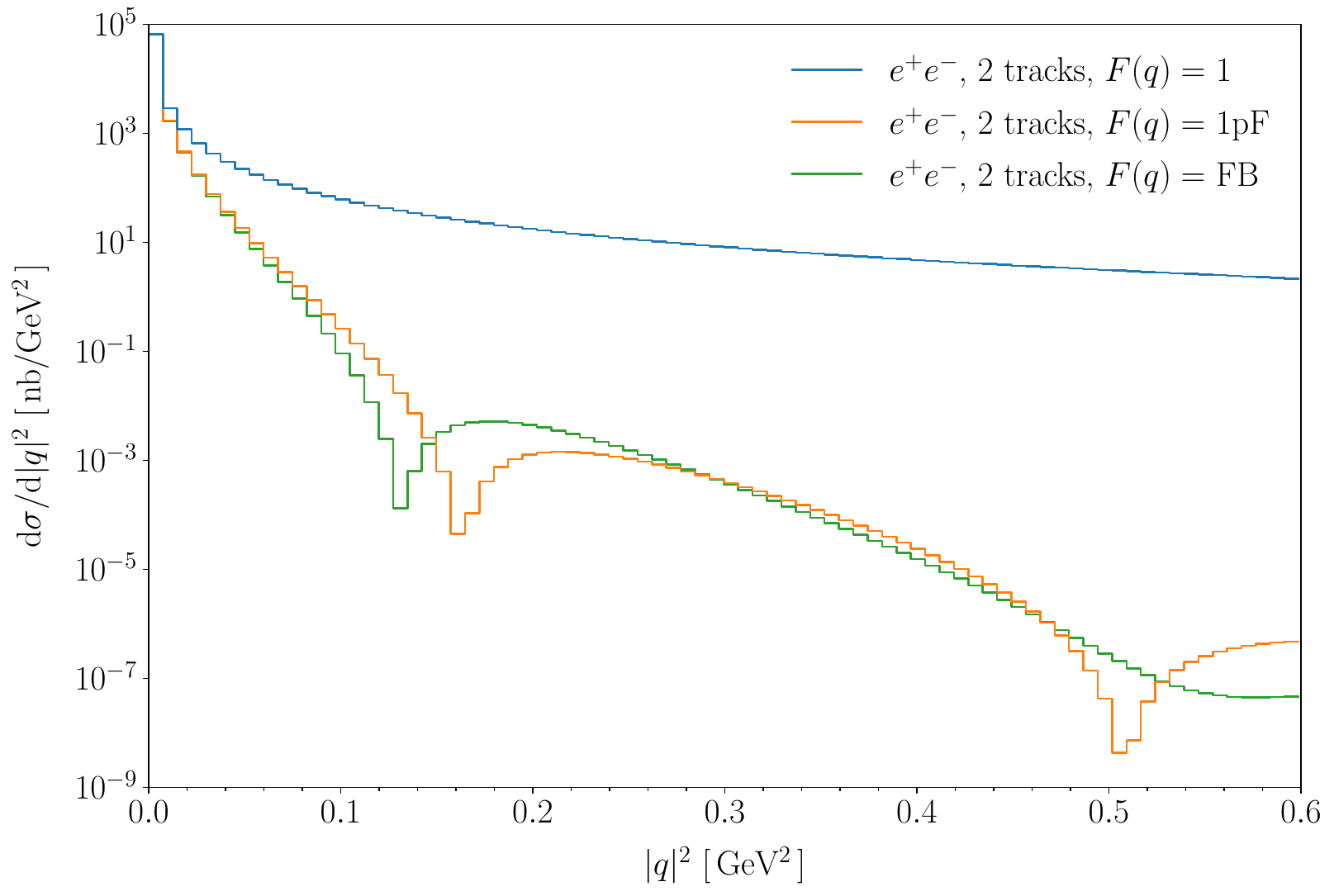}
    \caption{Differential cross section with respect to the squared exchanged momentum $\vert q \vert^2$ for two-track events and different form factor models.}
    \label{fig:q2_transf}
\end{figure}

In order to understand the effect of the nuclear form factor on differential distributions, we display in Fig.~\ref{fig:q2_transf} the differential distribution 
$\D \sigma / \D \vert q\vert^2$ for the process ${\mu^+ \,\text{C} \to \mu^+ \,\text{C}\: e^+ e^-}$, where $q^2$ is the squared space-like four-momentum that is exchanged with the nucleus field,  $q = p_1 - p_2 - p_3 - p_4$. Two different form factor models, 
1pF of Eq.~(\ref{eq:1pF}) and FB of Eq.~(\ref{eq:FB}), are considered and compared with the prediction obtained without form factor. The damping effect of the form factor for increasing $\vert q \vert^2$ and the typical oscillating behaviour are clearly visible. 
In the limit as $\vert q \vert \to 0$ the impact of the form factor vanishes, 
as expected. 
\begin{figure}[t]
    \centering
    \includegraphics[width = \columnwidth]{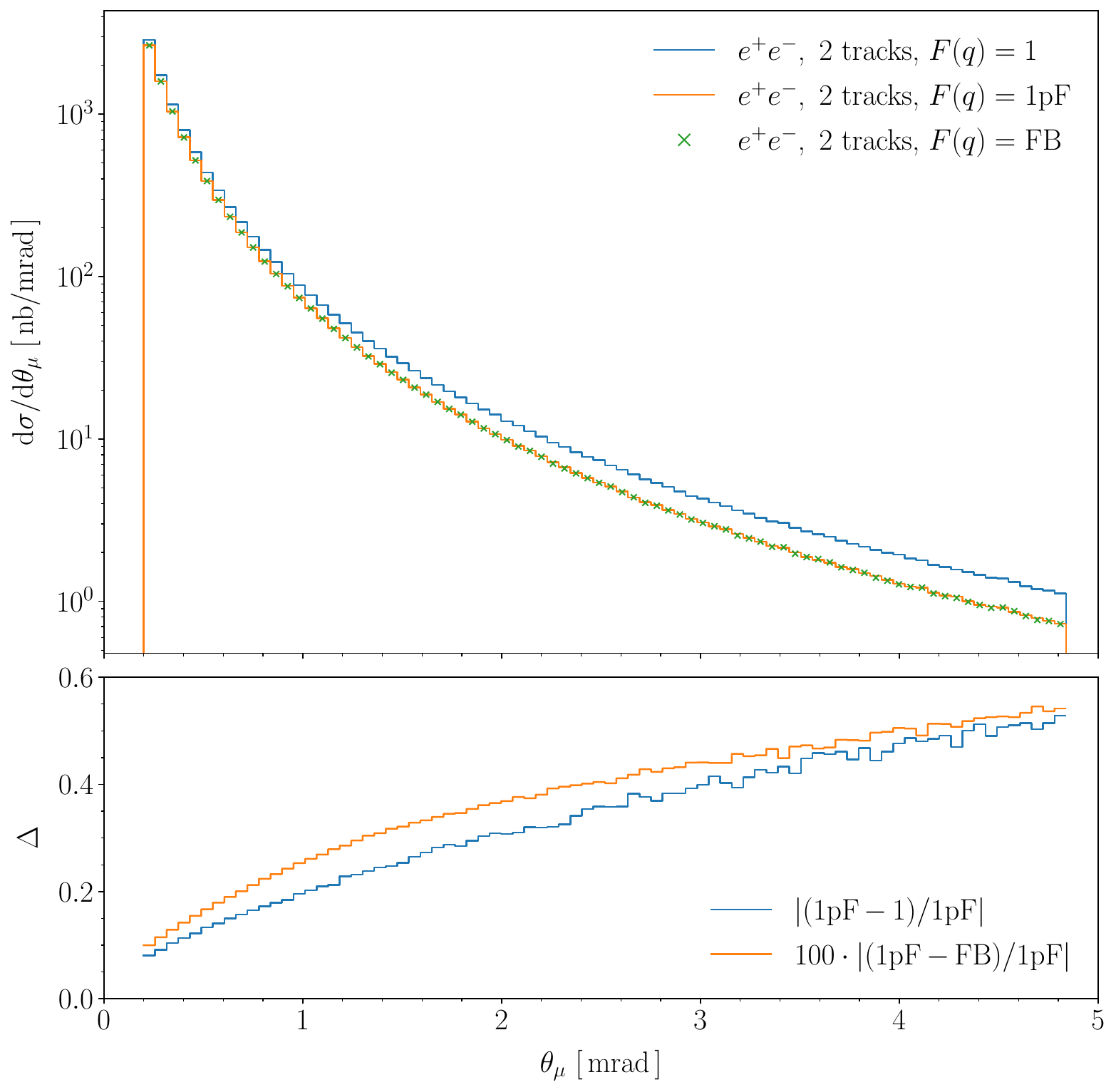}
    \caption{Differential cross section with respect to the muon angle $\vartheta_\mu$ for two-track events and different form factor models. The relative differences between the different cases are reported in the lower panel.}
    \label{fig:th_mu_ff}
\end{figure}
Fig.~\ref{fig:th_mu_ff} shows the distribution $\D \sigma / \D \theta_\mu$ for the process ${\mu^+ \,\text{C} \to \mu^+ \,\text{C}\: e^+ e^-}$ with exactly two tracks in acceptance. The blue solid line is obtained without introducing the form factor while the orange and green lines correspond to the 1pF and FB models, respectively. As can be seen from the lower panel, the relative effects of including the form factor range  monotonically from about 10\% for $\theta_\mu \sim 0.2$~mrad to about 50\% for $\theta_\mu \sim 4.8$~mrad. The relative differences between the predictions obtained with the two form factor models (see the orange line in the lower panel) are lower than 0.6\% in the whole relevant angular range. 

\subsection{Impact of acoplanarity and elasticity cuts}
\label{subsec:num-aco-el-cuts}
\begin{figure}[t]
    \centering
    \includegraphics[width = \columnwidth]{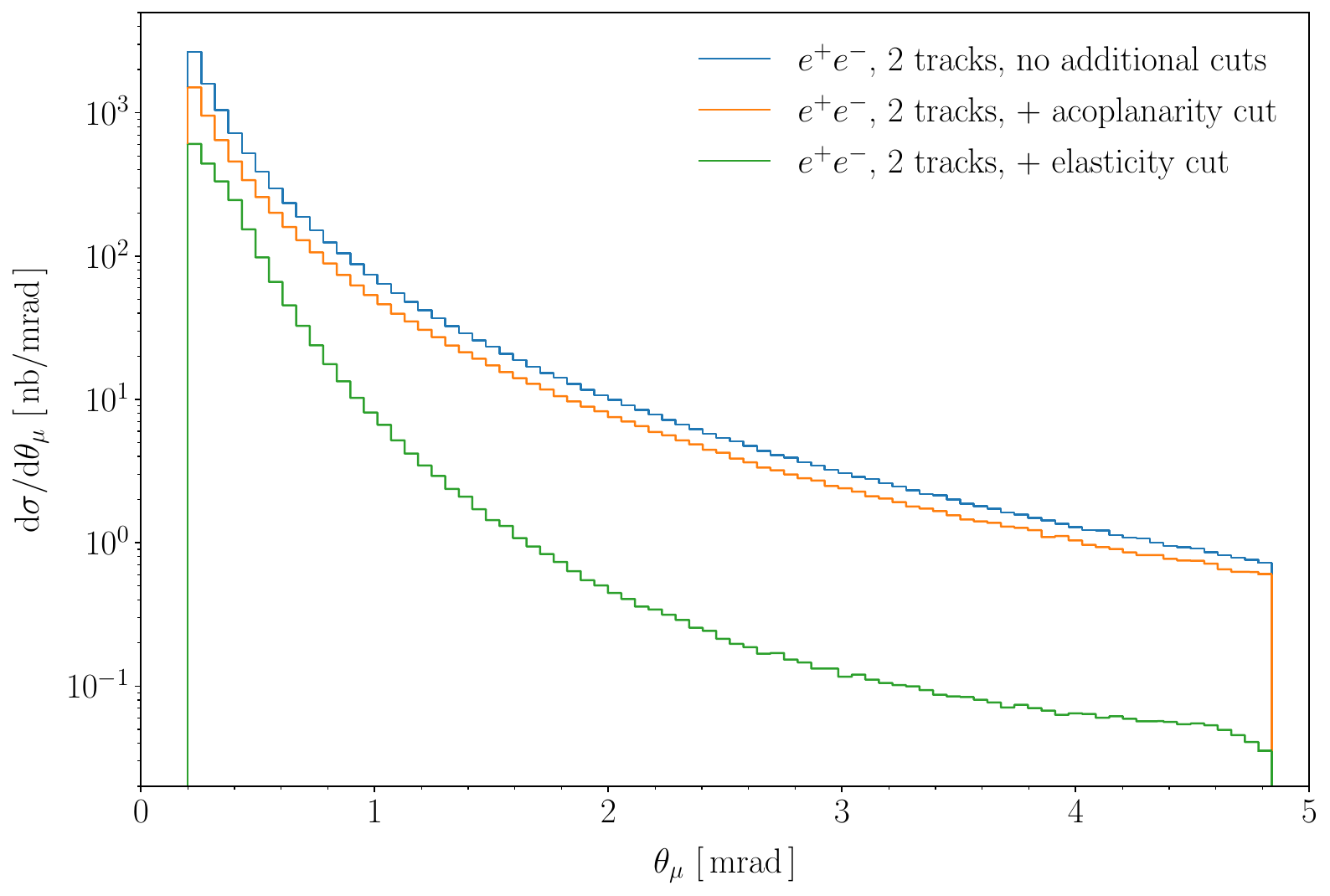}
    \caption{Differential cross section with respect to the muon angle $\vartheta_\mu$ for two-track events and different event selections.}
    \label{fig:th_mu_cut}
\end{figure}

In this subsection, the effect of the acoplanarity cut ${\xi < \xi_c = 400}$~mrad and the elasticity cut ${\delta < 0.2}$~mrad is discussed for the process ${\mu^+ \,\text{C} \to \mu^+ \,\text{C}\: e^+ e^-}$ with two reconstructed tracks, using the 1pF model as nuclear form factor.
This is the most relevant case for MUonE, because it has the same signature of an elastic event in terms of detected particles.
Fig.~\ref{fig:th_mu_cut} shows the separate impact of the acoplanarity and elasticity cuts on the differential distribution $\D \sigma / \D \theta_\mu$. While the former leads to a decrease of the cross section by an amount of less than a factor of two, the latter has a stronger impact, up to a factor of ten in the reduction of the cross section for larger muon scattering angles. 

\begin{figure}[t]
    \centering
    \includegraphics[width = \columnwidth]{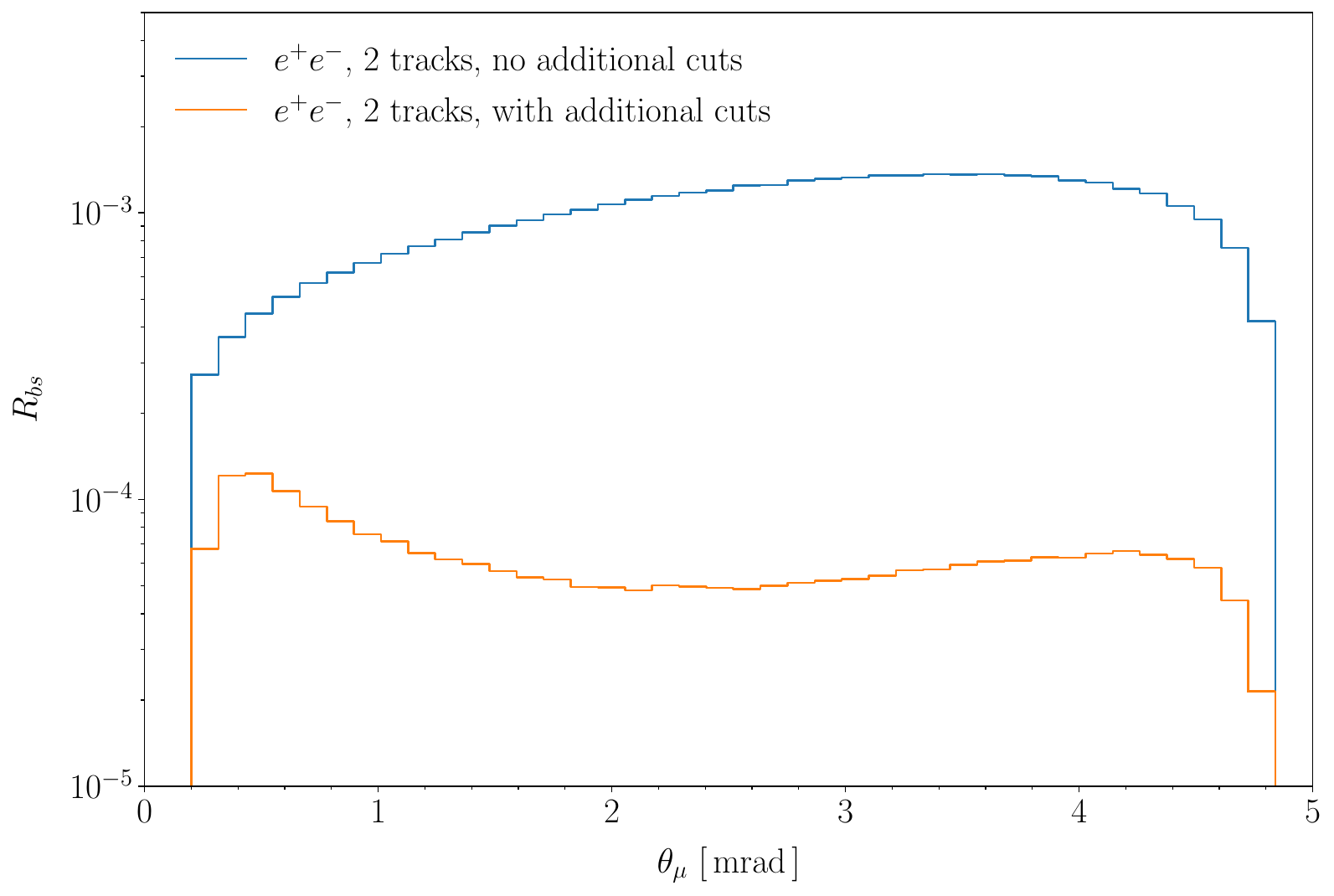}
    \caption{The background-to-signal ratio $R_{bs}$ as a function of the muon angle $\vartheta_\mu$ for different event selections.}
    \label{fig:th_mu_ratio}
\end{figure}

Finally, in order to have a rough estimate 
of the impact of the process ${\mu^+ \,\text{C} \to \mu^+ \,\text{C}\: e^+ e^-}$ on MUonE observables, we study in Fig.~\ref{fig:th_mu_ratio} the 
background-to-signal ratio as a function of the muon scattering angle, which is equal to the following differential cross section ratio: 
\begin{equation}\label{eq:delta_BS}
    R_{bs} = \frac{\frac{d\sigma}{d\theta_\mu}\left(\mu^+ X \to \mu^+ X e^+ e^-\right)}{Z\times\frac{d\sigma}{d\theta_\mu}\left(\mu^+e^-\to\mu^+e^-\right)} \,.
\end{equation}

The cross section for the elastic $\mu e$ 
process is calculated with \textsc{Mesmer} with QED radiative corrections at the next-to-next-to-leading order~\cite{Alacevich:2018vez,CarloniCalame:2020yoz,Budassi:2021twh}. 
The $Z$ factor takes into account that the nucleus is surrounded by $Z$ electrons. 

The blue line, which corresponds to the event selection with \textit{basic acceptance cuts}, shows that $R_{bs}$ ranges from few $10^{-4}$ to about $0.1\%$. The introduction of acoplanarity and elasticity cuts reduces the ratio by about one order of magnitude. This is a nice feature because a conservative uncertainty on the form factor of the order of $10$\% would allow a solid background subtraction with a total uncertainty below the $10^{-5}$ level.
The differences below the percent level between two form factor models, as shown in Fig.~\ref{fig:th_mu_ff}, show that a reliable background subtraction is feasible, in view of the MUonE target precision.  
Using a beryllium target, the ratio $R_{bs}$ would be further reduced by a factor of~$6/4$.

Since, in general, the QED corrections are 
expected to give contributions of the order of few percents, with possible enhancements in particular kinematic regions, their calculation at next-to-leading order could be necessary for the final analysis of the MUonE data. 

%=========================================%

\section{Conclusions}
\label{sec:concl}

In this work, we present a new fully differential calculation for the production of a real lepton pair from the scattering of an incoming muon on a nucleus at rest, namely $\mu^\pm X \to \mu^\pm X \ell^+ \ell^-$, where $\ell=\{e,\, \mu\}$. This process is the most relevant 
background for the MUonE experiment, which aims to perform a very high-precision measurement of the 
$\mu e$ scattering process. The precision requirements of MUonE make unreliable the 
simulation of the process through the \textsc{Geant4} toolkit, where, for instance, the outgoing muon scattering angle is neglected. Starting from the exact phase space, with 
energy conservation within the leptonic system, i.e. treating the nucleus as an external 
electromagnetic field, we consider the complete tree-level matrix element, including 
all diagrams and finite fermion mass effects. The finite extension of the nucleus is described through a 
nuclear form factor, with different models (and related parameters) taken from the 
existing literature. The calculation has been implemented in the Monte Carlo event 
generator \textsc{Mesmer}, available for detailed simulations of the MUonE experiment.

We presented a collection of numerical results for typical 
event selections currently used for MUonE simulations. 
We find that the production of $\mu^+ \mu^-$ pairs remains below the MUonE precision 
target of $10^{-5}$ with respect to the elastic signal, while the production of 
$e^+e^-$ pairs can be potentially dangerous. In particular, with \textit{basic acceptance cuts}, the background cross section can be of the order of $10^{-3}$ times the signal cross section. The acoplanarity and elasticity cuts are crucial for the reduction of the  background-to-signal ratio at the $10^{-4}$  level. 
Since the comparison of different nuclear form factor models shows an agreement on differential distributions below the percent level, a reliable background subtraction through Monte Carlo simulation is feasible for the MUonE precision requirement. As an independent handle, the fact that the differential cross sections for events with three tracks are of the same order of the irreducible two-track signatures allows to perform independent cross-checks on the two-track background estimate by measuring the three-track cross sections multiplied by the calculation of the ratio of two-track over three-track differential cross sections. In this ratio, the uncertainties on the nuclear form factor tend to cancel. 

As a last comment on the phenomenological results discussed in the present study, we stress that they should be  properly quantified by means of detailed analysis involving also detector simulation and track reconstruction, which will be carried out within the MUonE experimental collaboration. 

In view of the relevance of the considered process as a background to the high-precision differential measurements of MUonE, also QED corrections to the tree-level approximation should be considered, for a fully reliable estimate of the numerical 
impact of the process. In fact, such corrections are expected to give 
effects on the shape of the distributions at the few per cent level. 
This is left for a future investigation. 

\section*{Acknowledgements}

We are sincerely grateful to all our MUonE colleagues for the pleasant and exciting atmosphere and collaboration, which build the framework of the present study. 
We are particularly indebted to Guido Montagna for the useful discussions and the careful reading of the manuscript.
We warmly thank Vladimir Ivantchenko for the expert help on the use of \textsc{Geant4} and many useful discussions
on its current physics content and possible improvements.

This work is partially supported by the Italian Mini-stero dell'Universit\`a e Ricerca (MUR) and European Union - Next Generation EU through the research grant number 20225X52RA ``MUS4GM2: Muon Scattering for $g-2$'' under the program PRIN 2022. 

\bibliographystyle{elsarticle-num} 
\bibliography{nuclpair}

\end{document}